\documentclass[twocolumn,showpacs,prl]{revtex4}
\usepackage{graphicx}
\usepackage{amsmath}
\begin{document}

\title{Coarse-grained probabilistic automata
       mimicking chaotic systems}

\author{Massimo Falcioni}
\affiliation{ Dipartimento di Fisica Universit\`a di Roma
  ''La Sapienza''\\
 INFM, Unit\`a di Roma1 and SMC Center \\
 p.le Aldo Moro 2, 00185 Roma, Italy}

\author{Giorgio Mantica}
\email{giorgio.mantica@uninsubria.it} \affiliation{Center for
Nonlinear and Complex Systems, Universit\`a dell'Insubria, Via
Valleggio 11, 22100 Como, Italy, \\ and I.N.F.M., Unit\`a di Como,
I.N.F.N. Sez. Milano}

\author{Simone Pigolotti}
\affiliation{SISSA, Via Beirut 4, 34014 Trieste, Italy}

\author{Angelo Vulpiani}
\affiliation{ Dipartimento di Fisica Universit\`a di Roma
  ''La Sapienza''\\
 INFM, Unit\`a di Roma1 and SMC Center \\
 p.le Aldo Moro 2, 00185 Roma, Italy}

\begin{abstract}
Discretization of phase-space usually nullifies chaos in dynamical
systems. We show that if randomness is associated with
discretization dynamical chaos may survive, and be
indistinguishable from that of the original chaotic system, when
an entropic, coarse-grained analysis is performed. Relevance of
this phenomenon to the problem of quantum chaos is discussed.
\end{abstract}


\maketitle

Chaos is a common characteristic of motion in continuous spaces.
It is signaled by many indicators. Among these, positivity of the
Kolmogorov-Sinai (K-S) entropy is perhaps the most significant,
both in theory and applications \cite{avez,poli}. On the other
hand, trajectories in {\em discrete} spaces are always asymptotically
periodic---hence, of null K-S entropy. They may arise in the
discretization of continuous systems, as in the numerical
simulation of differential equations, but arguably their r\^ole is
most significant in the correspondence of classical and quantum
dynamics.

Consider for instance the paradigmatic example of the quantum
Arnol'd cat. Its dynamics are algorithmically equivalent to
classical motion on a regular lattice, whose spacing is inversely
proportional to the Planck constant \cite{berry,joegio}. When the
spacing diminishes, the lattice becomes denser in continuous,
classical phase-space. Yet, it has long been recognized that chaos
can{\em not} be naively revived in such a limit procedure
\cite{joegio,noi1,viva}. A way out of this {\em impasse} is
obtained by randomly perturbing the dynamics \cite{ott}: is this
addition enough to bring back the full algorithmic content, that
is the distinctive signature of chaos \cite{joe} ? We attempt here
an answer to this question in the most general and simple terms.

Deterministic-probabilistic systems (such as those occurring in
cellular automata \cite{grass}) have long been investigated, with
respect to their invariant measures \cite{ruel}, and also to the
entropic content of their motion \cite{karol}. We now extend this
study to the regime where continuous, discrete, and random effects
are simultaneously present, and intermix in non-trivial ways.

 Let us therefore consider the deterministic map
\begin{equation}
\label{eq:map} x_{t+1}=f(x_t),
\end{equation}
where $t$ is discrete time, $x$ belongs to $[0,1]^D$ and $D$ is
the dimension of the space. We embed in $[0,1]^D$ a uniform
rectangular lattice, of spacing $\eta$, and we label its states by
integer vectors $n$ in $[1, \lfloor \frac{1}{\eta} \rfloor ]^D$
($\lfloor \cdot \rfloor$ is the integer part) \cite{notadis}.  We
then restrict the map $f$ to act on this lattice, and we add
randomness:
\begin{equation}
\label{eq:discmap}
 n_{t+1}=  \lfloor \frac{1}{\eta} f( \eta \,
n_t) \rfloor  + \sigma_t \, .
\end{equation}
We stipulate that
the uncorrelated, random ``jumps'' $\sigma_t$,
extend to lattice neighbors with total probability $p$, so
that $\sigma_t = 0$ with probability $1-p$. Note that when $p=0$
the system is purely deterministic.
For definiteness, we shall study the generalized tent map in one
dimension:
\begin{equation}
x_{t+1} = \left\{
\begin{array}{lcl}
 {a x_{t}} \quad  \text{for} \quad 0 \leq x \leq  \frac {1}{a} & \\
 a(1 - x_{t})/(a-1) \quad \text{for} \quad  \frac {1}{a} \leq x \leq 1 &
\end{array}
\right. \label{eq:tent}
\end{equation}
with $a=3$, the two-dimensional Arnol'd cat map \cite{avez}:
\begin{equation}
\begin{array}{lll}
x_{t+1} =   x_{t} + y_{t} \;  \quad \text{mod} \;\; 1, &
\\ y_{t+1} =  x_{t} + 2 y_{t} \quad \text{mod} \;\; 1 &,
\end{array}
\label{eq:cat}
\end{equation}
and their probabilistic lattice automata, eq. (\ref{eq:discmap}).

We are now ready to briefly introduce our analytical tools. Let
$\{E_1,E_2,\ldots \}$ be a finite partition of phase space
consisting of identical hyper-cubic cells of side $\epsilon$. Let
$w_\epsilon=w_\epsilon^1, w_\epsilon^2, \ldots, w_\epsilon^n$ be a
finite symbolic trajectory, of length $n=|w_\epsilon|$:
$w_\epsilon^i = j$ if and only if $x_i \in E_j$. Let also
$p(w_\epsilon)$ be the frequency of $w_\epsilon$, defined by the
physical ergodic measure. The $n$-block entropies $H_n(\epsilon)$,
 $n=1,2,\ldots$,
are defined by the sums
\begin{equation}
\label{eq:cacca}
       H_n(\epsilon) = - \sum_{w_\epsilon  \;
      :  \;|w_\epsilon| = n}
p(w_\epsilon) \log p(w_\epsilon).
\end{equation}
The {\em partition entropy} $h(\epsilon)$ is the limit of
$H_n(\epsilon)/n$, or, with faster convergence, of the {\em
information rate} $h_n(\epsilon)= H_n(\epsilon) -
H_{n-1}(\epsilon)$, as $n \rightarrow \infty$ \cite{shan,koleps}.
The K-S entropy $h_K$ is the supremum of the partition-entropies,
with respect to all countable partitions; hence, $h(\epsilon) \le
h_K$, and $h_K$ can be obtained letting $\epsilon$ go to zero.
However, the function $h(\epsilon)$ for finite $\epsilon$ is
interesting in its own right \cite{gasp}. This is because it
gauges the rate of information production, for observations of
finite accuracy, as a function of the resolution desired. In line
with our approach in \cite{noi1} we also ascribe importance to the
full behavior of $H_n(\epsilon)$ versus $n$, and not only to the
limit $h(\epsilon)$.

It is well-known that the systems (\ref{eq:tent}), (\ref{eq:cat})
have positive K-S entropy. In contrast, the dynamics of the purely
discrete systems ({\em i.e.} eq. (\ref{eq:discmap}) with $p=0$)
are periodic, hence of null entropy. Yet, at scales larger than
the lattice spacing, $\epsilon > \epsilon_m := \eta$, they
approximate the continuous dynamics for a finite time, roughly of
the order of the logarithm of the period of the trajectory
\cite{rannou,disfranco,levy,coste,grebogi}. As a consequence, the
entropies $H_n(\epsilon)$ are also close to those of the
continuous system, for $n \le \overline{n}$. The upper boundary
$\overline{n}$ can be estimated requiring that at $n=\overline{n}$
the number of different $\epsilon$-histories of length $n$ of the
continuous system, ${\cal N}_{\epsilon} (n) \sim \exp (h(\epsilon)
\, n)$, be of the same order of the number of discrete states, $ M
\sim (1/ \eta)^{D}$. Here $D$ is the dimension of the attractor
or, in the absence of this latter, the dimension of the space.
This leads to:
\begin{equation}
\overline{n} \sim \frac{\log { M}}  {h(\epsilon)} \,
 \sim - \frac{ D \log \eta} {h(\epsilon)}.
\label{form1}
\end{equation}

Dependence of $\overline{n}$ on the average period of
trajectories, $T$, follows equally well.  Since ${ T} \sim
M^{D_2/2}$, where $D_2$ is the correlation dimension of the
ergodic measure \cite{grebogi},
\begin{equation}
\overline{n} \sim \frac{2 D}{D_2} \frac{\log { T}}{h(\epsilon)} .
\label{form2}
\end{equation}

In the deterministic discrete systems (eq. (\ref{eq:discmap}) with
$p=0$), $\overline{n}$ may be large enough to observe the entropic
growth of the continuous map, and sometimes to compute the entropy
$h(\epsilon)$ of the latter to a fair accuracy.  However, when $n$
exceeds $\overline{n}$, the $n$-words entropies, $H_n$, quickly
saturate at a constant value, of the order of $ \log { T} $, or
$\log (1/\eta)$, or $\log M$, which reveals the periodic regime of
the dynamics. It is clear that while the time $\overline{n}$ is
model-dependent, its logarithmic scaling with the parameters is
universal: the discrete, ``pseudo'' chaos seems to be very
short-lived \cite{joegio,noi1,viva}. It is at this point that the
random jumps $\sigma_t$ completely change the scenario.

First of all, the null-entropy, periodic system is turned into an
aperiodic stochastic process of maximal entropy $h_{p}$. If
$\sigma_t$ extends to $k$ neighbors with equal probability $p/k$,
then $h_{p} = -p \ln (p/k) - (1-p) \ln (1-p)$. Since $h_{p}$
totally originates from the random jumps at the lattice scale
$\epsilon_m$, it can be fully detected only at this scale: $h_{p}
\simeq h(\epsilon _m) \geq h(\epsilon)$, for $\epsilon \ge
\epsilon_m$.

It is clear that $h_p$ has no relation with the K-S entropy $h_K$.
Then, two cases must be considered. In view of the above, when
$h_{p} < h_K$, {\em addition of randomness, and observational
coarse-graining, cannot achieve the full entropy content of the
continuous system}. This no-go rule does not apply in the opposite
case. We find that {\em when $h_{p} \gg h_K$ and $\epsilon \gg
\epsilon_m$, the partition-entropies of the continuous system and
of the discrete ones tend to coincide}. Notice that the condition
$\epsilon \gg \epsilon_m$ requires that the number of discrete
states per cell, $M \, \epsilon^D$, be large.

\begin{figure}
\includegraphics[scale=.7]{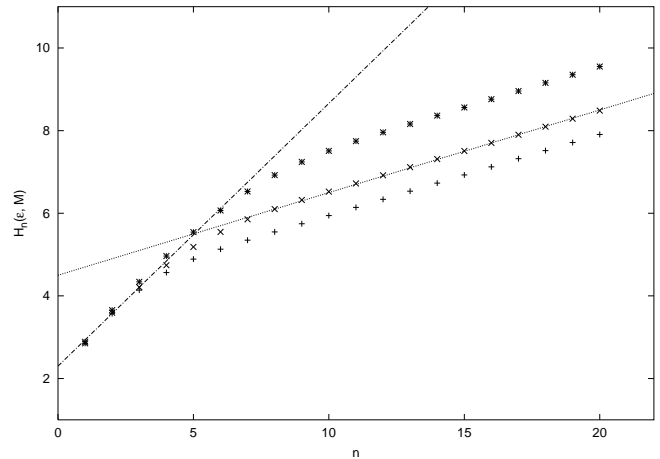}
\caption{\label{h-low} $n$-word entropies $H_n(\epsilon,M)$ vs.
$n$ for the discretized tent map with $\epsilon=1/18$, and $M =
252$ ($+$), $1008$ ($\times$), and $4032$. $H_n(\epsilon,M)$ are
compared to the lines with slope $h(1/18) \simeq 0.20$ and $h_K
\simeq 0.636$. }
\end{figure}

To prove these claims, let us first consider the tent map, eqs.
(\ref{eq:discmap},\ref{eq:tent}), with $p=0.05$, and
nearest-neighbor random jumps. In this case $h_p \simeq 0.233$ is
much smaller than $h_K \simeq 0.636$. Fig.~\ref{h-low} plots
$H_n(\epsilon,M)$ versus $n$, for different values of $M$, and
$\epsilon = 1/18$. This partition is generating, so that the
partition entropy of the continuous map, $h_{\text{cont}} (1/18)$,
is equal to $h_K$. We observe that for $n < \overline{n}$,
$h_n(\epsilon)$ is approximately equal to $h_K$, the entropy of
the continuous system. Later on, for $n
> \overline{n}$, the curve $H_n$ bends and---rather than tending to
a constant, as it would if randomness were not present---redirects
its growth to a different linear regime: $H_n \simeq h(\epsilon,M)
(n-\overline{n}) \, + A \log M$, with $A$ a suitable constant.
Numerically, we also find that $h(\epsilon,M) \simeq 0.20 \approx
h_p$, independently of $M$, even if $\epsilon= 1/18$ is much
larger than $\epsilon_m$. This is noteworthy: were the evolution
driven only by the probabilistic diffusion, $n_{t+1}=
 n_t + \sigma_t $, the $\epsilon$-entropy would have been ten times
smaller, $h_{\text{diff}}(1/18,252) \simeq 0.02$. The effect of
randomness is strongly enhanced by the deterministic evolution
\cite{berger}.

\begin{figure}
\includegraphics[scale=.7]{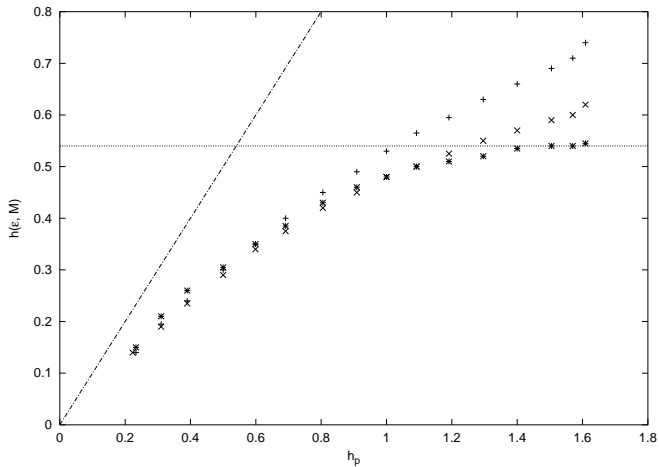}
\caption{\label{h-di-hp} Partition-entropies of
the discretized tent map $h(\epsilon,M)$ vs.  $h_p$,
with $k=4$, $\epsilon=1/7$ and $M = 63$ ($+$), $M = 168$ ($\times$)
and $M = 4032$. The dotted line is drawn at $h_{\text{cont}}(1/7)$,
the partition-entropy of the continuous system for $\epsilon=1/7$. }
\end{figure}

By raising the value of the jump probability $p$ \cite{noise} the
entropy $h_p$ increases, and it may exceed the K-S entropy $h_K$,
or $h_{\text{cont}}(\epsilon)$ for a given $\epsilon$.
Fig.~\ref{h-di-hp} plots $h(\epsilon,M)$ versus $h_p$, for
different values of $M$. At fixed, low $h_p$, the $M$ dependence
is rather mild, as was observed in Fig.~\ref{h-low}, and
$h(\epsilon,M)$ is smaller than $h_{\text{cont}}(\epsilon)$.
Raising $h_p$ at fixed $M$ to well exceed this value, and then
increasing $M$, we obtain convergence of $h(\epsilon ,M)$ to
$h_{\text{cont}}(\epsilon)$.

\begin{figure}
\includegraphics[scale=.7]{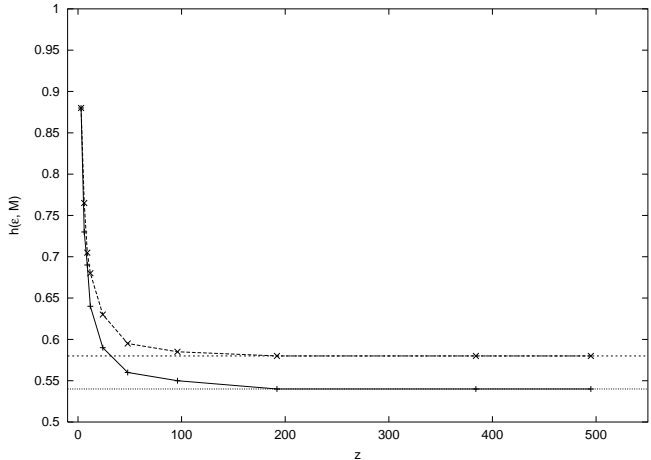}
\caption{\label{h-di-z} $\epsilon$-entropies $h(\epsilon,M)$
versus $z = M \epsilon$, for the discretized tent map, with
$\epsilon = 1/7$ and $\epsilon = 1/10$ ($\times$). Horizontal
lines are drawn at $h_{\text{cont}}(1/7 ) \approx .54$, and
$h_{\text{cont}}(1/10) \approx 0.58$. }
\end{figure}

This convergence is further illustrated  in Fig.~\ref{h-di-z} by
fixing $h_p = 1.5$, by choosing $\epsilon =1/7$ and $\epsilon
=1/10$, and by plotting $h(\epsilon,M)$ versus $z= M \epsilon$,
the number of lattice points enclosed in a cell of the partition.
The different curves tend to coincide for small $z$, where the
entropies overshoot: when coarse graining is too fine (too few
states in a cell) the direct action of randomness is dominating.
This effect fades in the opposite direction: randomness becomes a
germ that gets scale-amplified by the dynamics, and
$h(\epsilon,z/\epsilon)$ tends to $h_{\text{cont}}(\epsilon)$, the
partition-entropy of the continuous system.

\begin{figure}
\includegraphics[scale=.55]{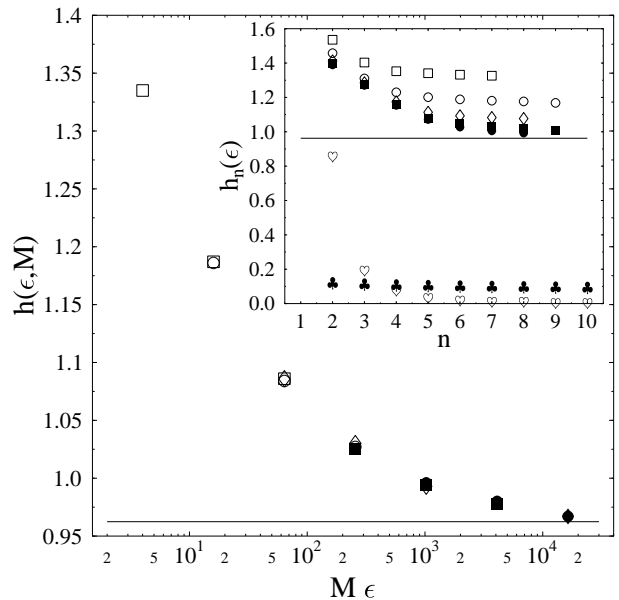}
\caption{\label{gatto} Entropies $h(\epsilon,M)$ vs. $M \epsilon$
in the perturbed Arnol'd cat map. Here, $M=16^2$ (squares), $32^2$
(circles), $64^2$ (diamonds), $128^2$ (filled squares), $256^2$
(filled circles), and $512^2$ (filled diamonds); $\epsilon=1/4,\,
1/16,\, 1/64$. The horizontal line is drawn at $h_K \simeq 0.962$.
In the inset, $h_n(\epsilon,M)$ vs. $n$ for $\epsilon=1/64$ and
$M=16^2, \ldots ,256^2$ (coded as before). At the largest value of
$M$, $M=256^2$, the purely discrete system (hearts) shows rapid
convergence to zero, in accordance with eqs.
(\ref{form1},\ref{form2}). The purely diffusive system (clubs) on
its part is consistently close to $h_{\text {diff}}(1/64,256^2)
\simeq .085$. }
\end{figure}

This behavior appears to be generic, as indicated by the results
for the two-dimensional Arnol'd cat map, subject to a random
perturbation with $k=4$ and $h_p=1.5$, plotted in
Fig.~\ref{gatto}. Since $\epsilon=1/4, 1/16, 1/64$, all provide
generating partitions, the corresponding values of
$h_{\text{cont}}(\epsilon)$ are equal to $h_K \simeq 0.962$.  As a
consequence the data fall on a single line, starting from the
entropy of the random perturbation, $h_p=1.5$, and converging to
the K-S entropy $h_K \simeq 0.962$. The convergence of
$h_n(\epsilon,M)$ to the asymptotic values is shown in the inset
of Fig.~\ref{gatto}, together with the partition entropies
obtained for the purely discrete system ($p=0$), and for the
purely stochastic motion.  This comparison proves that
$h(\epsilon,z/\epsilon)$ are truly asymptotic values, and provides
further evidence in support of our explanation of the phenomenon,
with which we now conclude.

Imposing a finite lattice to the otherwise continuum set of states
of a dynamical system, inevitably bounds the algorithmic
complexity of its trajectories, and the value of its partition
entropies, to the logarithm of the number of states. If the
lattice is perceived with some fuzziness---or if random errors are
allowed, following the approach of this paper---one expects that
on large scales continuum properties, and chaos might re-emerge,
and be indistinguishable from that of the original system
\cite{chorno}. We have determined the conditions for this to
happen. Firstly, observational coarse-graining must be invoked.
Secondly, the action of the external randomness must be confined
to the ``microscopic'', unresolved scales. The instability of
deterministic dynamics amplifies these microscopic, random errors,
and carries them over to the large, observation scales. Finally,
by a sort of conservation law, the {\em flow of information}
supplied by the microscopic ``zitterbewegung'' must not be less
than $h_K$, the maximum entropy production rate of the continuous
system.

For a long time, research in quantum chaos has looked for quantum
characteristics related to classical chaotic motion. The fact that
none of these could be properly called chaos led to the concept of
{\em pseudo-chaos}, and cast doubts on the very existence of chaos
in nature. The results presented in this paper suggest that one
might try to reverse this approach and consider classical dynamics
as an effective theory that, via truly chaotic deterministic
dynamical systems, models a randomly perturbed quantum motion
under observational coarse-graining. \hfill\break


G.M. was supported by
PRIN-2000 Chaos and localization in classical and quantum mechanics,
and by HPRN-CT-2000-00156 Quantum Transport on an atomic scale.
M.F and A.V. were supported by INFM (Centro di Ricerca e Sviluppo SMC)
and MIUR (Cofin. ``Fisica Statistica di Sistemi Complessi Classici
e Quantistici'').


\begin{thebibliography}{99}
\bibitem{notadis}
If the volume of the space accessible to the continuous system is
finite, uniform discretization results into a finite number of
states. It has been remarked that uniform discretization is
different from, and theoretically more appealing than,
floating-point discretization \cite{rannou},\cite{disfranco}.
\bibitem{noise}
Note that if $p \gg 1/{ T}$,
contamination of the random trajectory by any individual
periodic orbit of the discrete system is unlikely.
\end{thebibliography}
\end{document}